\pdfoutput=1
\documentclass[amsfonts,amssymb,amsmath,prl,aps,twocolumn,notitlepage,superscriptaddress,showpacs,floatfix,10pt]{revtex4-1}%
\usepackage[T1]{fontenc}
\usepackage{times}
\usepackage{xcolor}
\usepackage[colorlinks,linkcolor=blue,citecolor=blue]{hyperref}
\usepackage{graphicx}
\usepackage{bm}
\usepackage{dsfont}
\usepackage{amsmath}
\usepackage{amsfonts}
\usepackage{amssymb}%
\renewcommand{\vec}[1]{{\ensuremath{\bm{\mathrm{#1}}}}}

\begin{document}

\title{Thermal spin dynamics of yttrium iron garnet}
\author{Joseph Barker}
\affiliation{Institute for Materials Research, Tohoku University, Sendai 980-8577, Japan}

\author{Gerrit~E.W.~Bauer}
\affiliation{Institute for Materials Research, Tohoku University, Sendai 980-8577, Japan}
\affiliation{WPI-AIMR, Tohoku University, Sendai 980-8577, Japan} \affiliation{Kavli Institute of NanoScience, Delft University of Technology, Lorentzweg 1, 2628 CJ Delft, The Netherlands}

\begin{abstract}
Yttrium Iron Garnet is the prototypical material used to study pure spin
currents. It is a complex material with 20 magnetic atoms in the unit cell.
Almost all theories and experimental analysis approximates this complicated
material to a simple ferromagnet with a single spin wave mode. We use the
method of atomistic spin dynamics to study the temperature evolution of the
full 20 mode exchange spin wave spectrum. Our results show a strong frequency
dependence of the modes in quantitative agreement with neutron scattering
experiments. We find this causes in a reduction in the net spin pumping due to
the thermal occupation of optical modes with the opposite chirality to the FMR mode.

\end{abstract}
\maketitle

\textit{Introduction} -- Spin transport in magnetic insulators has attracted
much interest since the experimental demonstration of the Spin Seebeck effect
(SSE)~\cite{Uchida:2010ei}. The field of study is broadly termed spin
caloritronics and encompasses the coupling between spin, charge and heat
currents~\cite{Bauer:2012fq}. Typical experimental setups consist of bilayers
made of a ferrimagnetic insulator coated with a thin metallic film possessing
a large spin Hall angle. Of special interest is the ferrimagnetic insulator
Yttrium Iron Garnet (YIG) due to its very low damping, $\alpha\approx10^{-5}$
and therefore long-lived spin waves \cite{Cherepanov:1993dd,Sun:2012co}. The
magnetism is carried by localised Fe moments in 8 tetrahedral (minority) and
12 octahedral (majority) oxygen cages per unit cell, with an anti-parallel
ferrimagnetic state between the two coordinations. However, most recent
theories and experiments treat YIG as a \emph{ferro}-magnet with a single,
parabolic spin wave mode~\cite{Xiao:2010uc,Ritzmann:2014qn}, simply because
the influence of YIG's complex electronic and magnetic structure on spin
transport is not known. Ohnuma \textit{et al. }~\cite{Ohnuma:2013il}
introduced a simple two mode model to describe the basic aspects of
ferrimagnetic dynamics, but its spectrum bares little resemblance to that of
YIG~\cite{Cherepanov:1993dd}. In this Letter we show that the frequencies and
line widths of spin waves in YIG are strongly temperature dependent. We find
that at room temperature higher frequency spin wave modes are significant
occupied. Optical modes with opposite chirality to the acoustic mode turn out
to have a disproportionate effect on spin transport and must be taken into
account when modelling or interpreting, for example, the spin Seebeck effect.

Experimentally, techniques such as Brillouin light scattering give access to
the long wavelength, GHz frequency, dipole spin waves~\cite{Demokritov:2001kk}%
. Studying the THz frequency `exchange' spin wave modes requires expensive
inelastic neutron scattering experiments~\cite{Plant:1977tn}. The role of the
high frequency `thermal' magnons remains poorly understood, despite their
importance in interpreting recent
experiments~\cite{Kikkawa:2015bn,Jin:2015ik,Guo:2015vi,Cornelissen:2015cz,Geprags:2016hh}%
. Since spin wave spectra are material specific, improving the understanding
of general characteristics could aid in the selection of materials to progress
towards applications such as sensing and heat
scavenging~\cite{Uchida:2014bj,Uchida:2016jo}.

\textit{Atomistic model} -- YIG is an insulator with a large electronic band
gap. The $\mathrm{Fe}^{3+}$ ion d-shells are half-filled with spin value of
$S=5/2$ and can be modeled with the Heisenberg Hamiltonian
\begin{equation}
\mathcal{H}=-\frac{1}{2}\sum_{ij}J_{ij}\vec{S}_{i}\cdot\vec{S}_{j}-\sum_{i}\mu_{s,i}%
\vec{B}\cdot\vec{S}_{i}. \label{H}%
\end{equation}
Here $J_{ij}$ is the isotropic exchange energy between (normalized) spins with
$\left\vert \vec{S}_{i}\right\vert =1$, where the indices $i,j$ enumerate
sites on the YIG crystal lattice. The $\mathrm{Fe}^{3+}$ magnetic moment is
$\mu_{s}=g\sqrt{S(S+1)}{\mu_{\mathrm{B}}}=5.96{\mu_{\mathrm{B}}}$ (${\mu_{\mathrm{B}}}$ is the Bohr magneton). An external field $B_{z}=0.01$~T
defines the spin quantization ($z$) axis. The crystal magnetic anisotropy and
dipolar interaction of YIG are negligibly small compared to THz frequencies of
thermal spin waves and disregarded here. Neutron
scattering~\cite{Plant:1977tn} indicates that nearest neighbor exchange
dominates and we adopt $J_{ad}=-9.60{\times10^{-21}}$~J, $J_{dd}%
=-3.24{\times10^{-21}}$~J and $J_{aa}=-0.92{\times10^{-21}}$~J
~\cite{Cherepanov:1993dd}, where the subscripts refer to the majority ($d$)
and minority ($a$) spins. All couplings are antiferromagnetic, which in
principle could produce a frustrated state, but the dominating $J_{ad}$ is so
large that a perfectly anti-collinear ground state is stable. By Metropolis
Monte-Carlo calculations~\cite{Hinzke:1999ud} we compute the magnetization
$\vec{m}=\tfrac{1}{N_{a}}\sum_{i}\vec{S}_{a,i}+\tfrac{1}{N_{d}}\sum_{i}\vec
{S}_{d,i}$ as a function of temperature and find a Curie temperature of
$520$~K close to the experimental value of $559$~K~\cite{Anderson:1964kv} and
in agreement with other calculations~\cite{Oitmaa:2009jm}. Throughout this
work we address a bulk system by periodic boundary conditions for a supercell
of repeated unit cells.

The spin wave spectrum can be obtained by diagonalizing the Hamiltonian
(\ref{H})~\cite{Cherepanov:1993dd} but non-linear effects such magnon-magnon
interactions, thermal noise and damping require a different approach. At
finite temperatures, the spin moments fluctuate around the local equilibrium
state. The exchange coupling correlates the motion of all moments, giving rise
to collective spin waves. The dynamical structure factor (or spin wave power
spectrum) as measured by inelastic neutron scattering is the Fourier transform
of spatio-temporal spin-spin correlation functions. Here we compute the local
spin dynamics, and from them the structure factor, thereby avoiding a magnon
ansatz and including magnon-magnon interactions to all orders.

\begin{figure*}[ptb]
\includegraphics[]{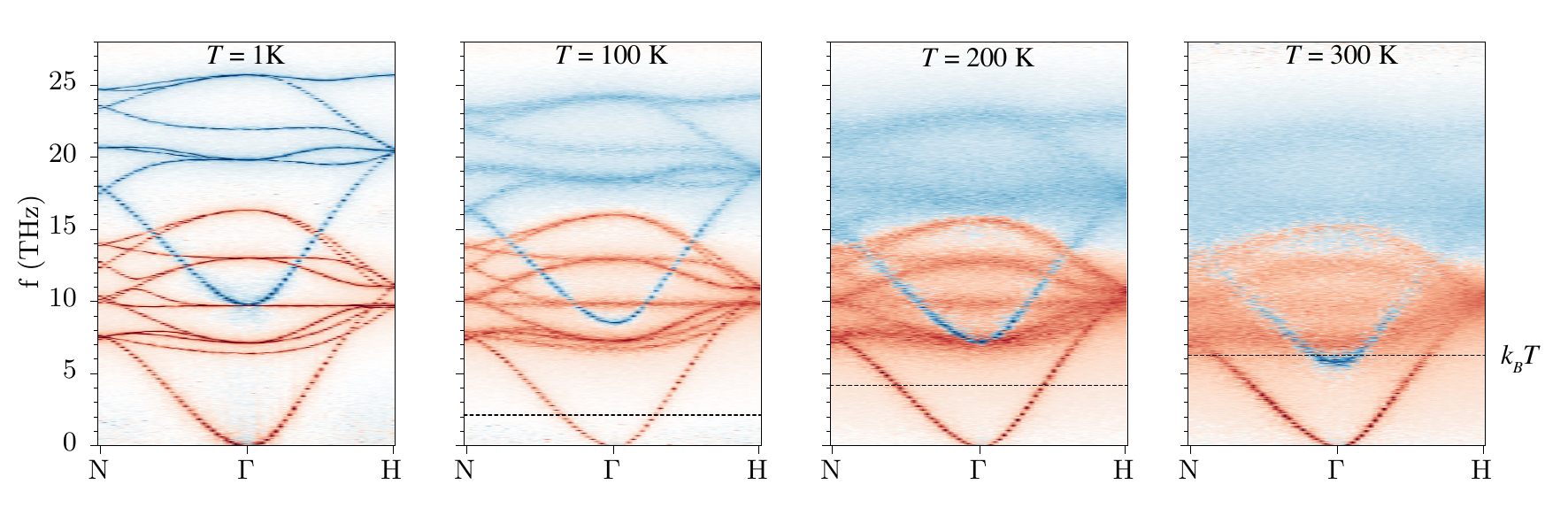} \caption{(Color online) YIG spin
wave spectrum as calculated for different temperatures. The dashed lines mark
$2\pi\hbar f={k_{\mathrm{B}}}T$. The red/blue coloring denotes the +/- mode
chirality relative to the magnetization direction.}%
\label{fig:spectrum}%
\end{figure*}

The spin dynamics is described by the atomistic Landau-Lifshitz-Gilbert
equation (LLG) based on the Hamiltonian (\ref{H})
\begin{equation}
\frac{\partial\vec{S}_{i}}{\partial t}=-\frac{|\gamma|}{(1+\lambda^{2}%
)\mu_{s,i}}(\vec{S}_{i}\times\vec{H}_{i}+\lambda\vec{S}_{i}\times\vec{S}%
_{i}\times\vec{H}_{i}). \label{eq:llg}%
\end{equation}
The local effective field is $\vec{H}_{i}=\vec{\xi}_{i}-\partial
\mathcal{H}/\partial\vec{S}_{i}$ where $\xi_{i}$ is a stochastic term with
\begin{equation}
\langle\xi_{i}(t)\rangle=0;\quad\langle\xi_{i}(t)\xi_{j}(t^{\prime}%
)\rangle=\delta_{ij}\delta(t-t^{\prime})2\lambda\mu_{s}{k_{\mathrm{B}}%
}T/\gamma.
\end{equation}
where $\langle\cdots\rangle$ denotes the statistical time average. The
gyromagnetic ratio is $\gamma=1.76{\times10^{11}}$ rad s$^{-1}$T$^{-1}$. The
measured Gilbert damping $\alpha$ of ferrimagnets is a combination of the
damping parameters of the sublattices. According to Wangsness' formula
$\alpha=(\lambda_{a}M_{a}+\lambda_{d}M_{d})/(M_{d}-M_{a})$%
~\cite{Wangsness:1953vga}. Here we use the damping parameter $\lambda
=\lambda_{a}=\lambda_{d}=2{\times10^{-5}}$ giving $\alpha={10^{-4}}$, which is
a typical (although not record) value. We solve the stochastic Langevin
equation (\ref{eq:llg}) with the Heun method, using a time step of $\Delta
t=0.1$ fs. The low damping requires careful equilibration. We first use a
Metropolis Monte-Carlo algorithm to converge to the equilibrium magnetization.
We then integrate the LLG dynamically for $1$~ns, which is sufficient to
achieve a steady state regime in the presence of noise. Finally we collect
data for $0.1$~ns which is Fourier transformed in space and time to distill
the spectral information.

In our classical formalism the thermal noise is white, and through
equipartition the system obeys Rayleigh-Jeans rather than Planck or
Bose-Einstein statistics for magnons. Our results at low temperatures and high
energies do not capture possible quantum effects. However, at elevated
temperatures quantum effects are suppressed by magnon-magnon interactions and
classical spin models can be expected to give a good agreement with experiments.

The spin wave spectrum is revealed in terms of structures in the space-time
Fourier transform of the spin fluctuations. The Fourier representation of the
spin dynamics reads
\begin{equation}
\mathcal{S}_{k}(\vec{q},\omega)=\frac{1}{\sqrt{2\pi}}\frac{1}{N_{c}}\sum
_{n=1}^{N_{c}}\sum_{\vec{r}}e^{\mathrm{i}{\vec{q}\cdot\left(  \vec{r}-\vec
{p}_{n}\right)  }}\int_{-\infty}^{+\infty}e^{{{\mathrm{i}}\omega t}}%
S_{k,n}(\vec{r},t)\mathrm{d}t
\end{equation}
where $\vec{p}_{n}$ is the position of the $n$-th spin (of a total of
$N_{c}=20)$ in the unit cell and $k=x,y,z$. The $q$-space resolution is
determined by the system size of $64\times64\times64$ unit cells (5,242,880 spins).

Thermal spin motive force is proportional to the transverse dynamical
susceptibility or equal-time spin correlation function~\cite{Xiao:2010uc}
which is related to the correlation function $\left\langle \dot{S}%
_{y,i}(0)S_{x,i}(0)-\dot{S}_{x,i}(0)S_{y,i}(0)\right\rangle $ that is
equivalent to the wave vector and frequency integral of the power spectrum
$\langle\omega\mathcal{S}_{x}(\vec{q},\omega)\mathcal{S}_{y}^{\ast}(\vec
{q},\omega)\rangle-\langle\omega\mathcal{S}_{y}(\vec{q},\omega)\mathcal{S}%
_{x}^{\ast}(\vec{q},\omega)\rangle.$ This correlation function is a Stoke's
parameter $V=-2\mathrm{Im}(\mathcal{S}_{x}\mathcal{S}_{y}^{\ast})$ and the
sign identifies the chirality or polarization of the
eigenvectors~\cite{Harris:1963vs}. The label `+' chirality implies
counter-clockwise rotation i.e. the precession direction of a magnetic moment
in an applied field e.g. under FMR.

\textit{Spin wave spectrum} -- In Fig.~\ref{fig:spectrum} we display the
calculated spin wave spectra for different temperatures. The coloring
indicates the chirality of the modes. Red modes have the `+' chirality, while
the blue modes precess in the opposite direction. The latter (optical) modes
are energetically costly due to the strong exchange field between the two
sublattices, so emerge only at frequencies above the exchange splitting.

At the lowest temperature considered (1~K) the amplitude of the excitations
(or number of magnons) is small and magnon interactions are very weak. The
calculated spectrum therefore agrees well with the linearized spin wave
theory~\cite{Cherepanov:1993dd}. The following discussion is focused on the
two nearly rigidly shifted parabolic modes with opposite chirality. The lowest
frequency mode is the ferromagnetic-like acoustic mode. The second mode is
blue shifted by a spin wave gap caused by the exchange field between the two
sublattices $\Delta=3J_{ad}\langle S_{z,a}\rangle-2J_{ad}\langle
S_{z,d}\rangle\approx10~$THz and is the optical, antiferromagnetic-like mode
between the two sublattices. We observe 5 additional flat modes in the 5 and
10~THz range that are thermally excited at room temperature. Since their mass
is very high, they are expected to only weakly contribute to spin transport
than the highly dispersive ones.

\begin{figure}[ptb]
\includegraphics[width=0.5\textwidth]{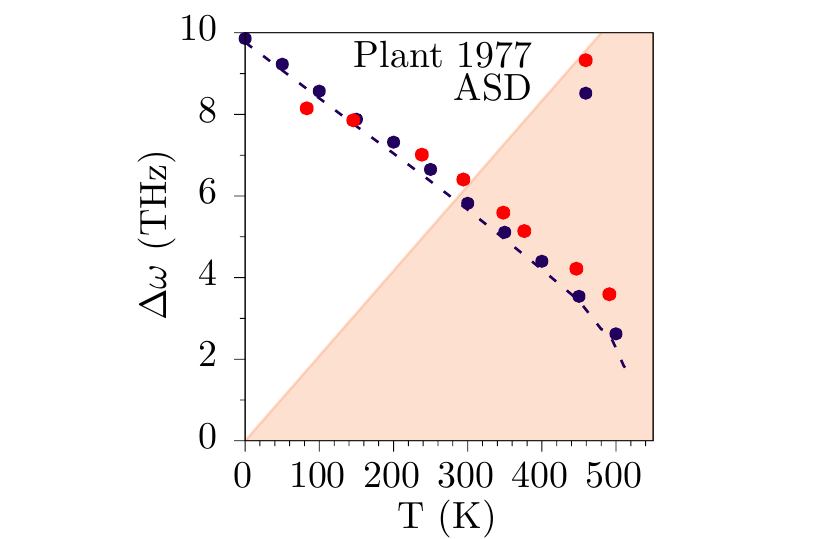} \caption{(color
online). Temperature dependence of the spin wave gap. Blue circles are the
calculations in this work, red points are the neutron scattering data of
Plant~\cite{Plant:1977tn}. The shaded area marks $\hbar\omega<{k_{\mathrm{B}}%
}T$. The dashed line is the reduction in exchange field due to spin
fluctuation.}%
\label{fig:yig_optical}%
\end{figure}

Thermal fluctuations reduce the magnetic order $\langle S_{z,a}\rangle$,
$\langle S_{z,d}\rangle$ and thereby the exchange field $\Delta$, as observed
in Fig.~\ref{fig:spectrum}. Our calculations agree very well with neutron
scattering experiments~\cite{Plant:1977tn} (Fig.~\ref{fig:yig_optical}). We
emphasize that our exchange constants $J_{ij}$ are not temperature dependent -
the effect is caused by statistical mechanics alone. Below $\hbar
\omega={k_{\mathrm{B}}}T$ (dashed line) spin waves modes are thermally
occupied. The occupation above this line is small and technically governed by
quantum statistics, but unimportant for (near to) equilibrium properties. Far
below room temperature only the ferromagnetic-like acoustic mode is
significantly occupied and the use of a single parabolic spin wave model is
justified. However, at room temperature and above, this approach breaks down
and the effects reported here must be taken into account in order to
understand the properties of YIG.

\begin{figure}[ptb]
\includegraphics[width=0.5\textwidth]{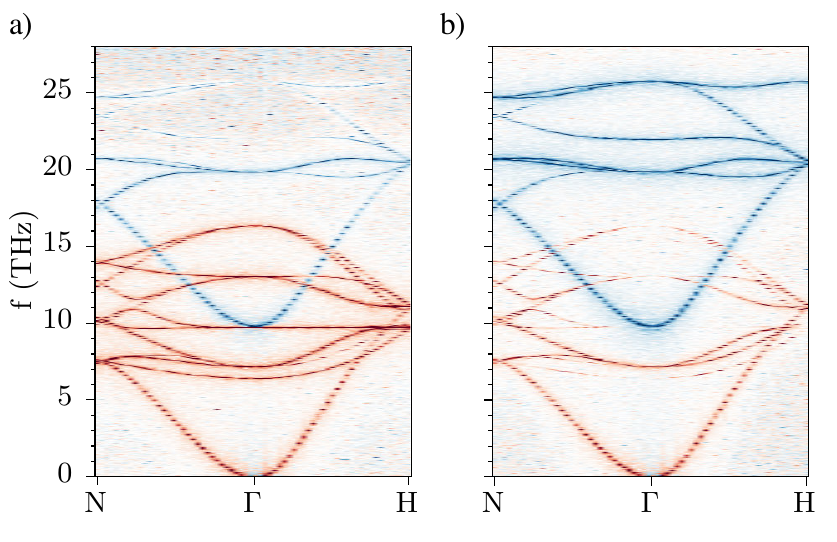} \caption{(color
online). Examples of the on-site correlations from different points in the
unit cell a) FeA at $(0,1/2,0)$ and b) FeA at $(1/2,0,0)$ in fractional
coordinates.}%
\label{fig:yig_dsf_unitcell}%
\end{figure}

Our method allows a comprehensive treatment of the non-linear thermodynamics
from low temperatures up to the magnetic phase transition. The magnon ansatz
often used to describe spin dynamics is based on the Holstein-Primakoff
transformation expanded to low order in the number of magnons. A larger
number of thermally excited magnons initially can be captured by magnon-magnon
interactions. These reduce the magnon lifetime as reflected by an increased
broadening of the spectral function. When the broadening becomes larger than
the spin wave energy splittings, the magnon concept breaks down completely.
This picture is illustrated by Fig.~\ref{fig:spectrum} in which we observe
that with increasing temperature the flat modes completely melt into an
incoherent background that can no longer be interpreted in terms of spin
waves. However, the coherence of the fundamental acoustic and optical modes
turns out to be remarkably robust against thermal fluctuations: the line
widths as well as the parabolic curvature, a metric of spin wave stiffness,
hardly change with temperature. Our formalism can provide information about
individual local moment fluctuations irrespective of the coherence of the spin
wave excitations, which can be used to shed light into this remarkable
behavior. In Fig.~\ref{fig:yig_dsf_unitcell} we plot the site-resolved
contributions to the power spectrum. We clearly see that the fundamental
acoustic and optic modes are homogeneously spread over the unit cell, while
the other modes are strongly localized on one of the local moments that are
consequently coupled by the small $J_{aa}$ and $J_{dd}$ exchange constants.
The spin waves with a flat dispersion are slack and susceptible to thermal agitation.

The resilience of both fundamental modes agrees with observations
\cite{LeCraw:1961ef} but appears to contradict common wisdom that the spin
waves stiffness decreases linearly with temperature~\cite{Bastardis:2012cs}.
The reason for the anomalous behaviour of YIG have not been well understood,
especially in the higher temperature regime where the optical magnons become
involved~\cite{Cherepanov:1993dd}. The present results indicate that the
decoupling of the fundamental extended modes from the floppy localized modes,
as discussed above, protects the spin wave stiffness as well as the lifetime.

\textit{Spin pumping} -- is the emission of spin currents from a magnetic
material into metal contacts by the magnetization dynamics. The latter can be
driven for example by microwaves and ferromagnet resonance or by thermal
excitations. In the presence of a temperature difference, the imbalance
between thermal spin pumping and spin transfer torque from the metal causes
net spin currents that can be converted into voltages by the inverse spin Hall
effect, i.e. the spin Seebeck effect. Spin pumping and SSE are interface
effects that are proportional to the spin-mixing interface conductance and
equal time and space\ spin correlation functions of the magnet at the
interface \cite{Xiao:2010uc}. When the interface correlations are identical to
that in the bulk, i.e. are not perturbed by the presence of the interfaces
(Golden Rule or tunneling approximation), our simulations provide direct
access to the spin Seebeck effect except for a constant that contains the spin
mixing conductance, spin Hall angle and temperature gradient. Our approach
does not address surface spin wave states or spin-pumping induced enhanced
damping. For samples much thicker than the magnon relaxation lengths thermal
gradients induce magnon spin and heat transport in the bulk of the
ferromagnet, which have been held responsible for e.g. YIG layer thickness
dependence of the SSE \cite{Kikkawa:2015bn,Guo:2015vi,Ritzmann:2014qn}. These
effects are beyond the scope of the present study, however.

\begin{figure}[ptb]
\includegraphics[width=0.48\textwidth]{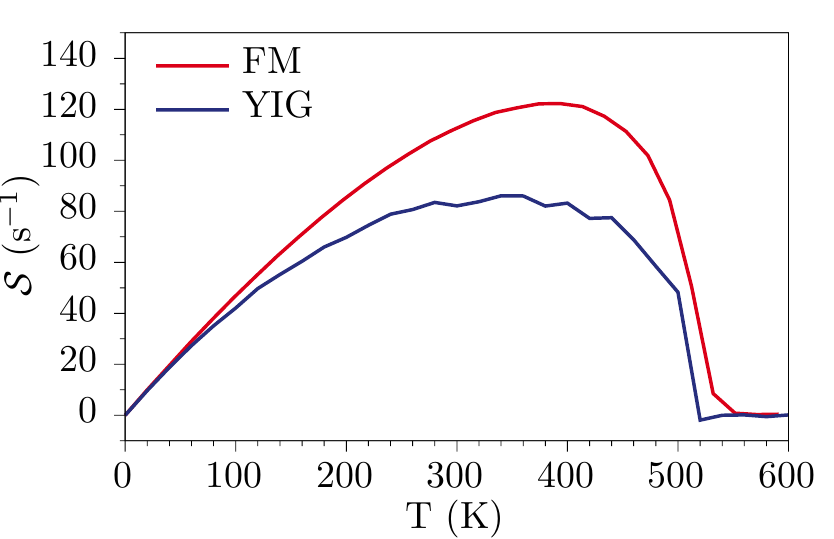} \caption{(color
online). Spin pumping (Eq. \ref{eq:spin_pumping})\ of YIG and a hypothetical
ferromagnet into a metal contact as a function of temperature.}%
\label{fig:spin_pumping}%
\end{figure}

The spin mixing conductance of an interface between a metal and a magnetic
insulator with local moments is governed by the exchange integrals between the
local moments and the conduction electrons \cite{Jia:2011gm}. The DC pumped
spin current then reads
\begin{equation}
\langle\vec{I}\rangle_{A}=\frac{\hbar}{4\pi}\frac{\operatorname{Re}%
g^{\uparrow\downarrow}}{N_{A}}\sum_{i}^{N_{A}}\langle\vec{S}_{i}\times
\dot{\vec{S}}_{i}\rangle
\end{equation}
where $N_{A}$ are the number of moments at the interface $A$. Averaging over
the (weak) dependence of the mixing conductance on the specific interface
\cite{Jia:2011gm} and to leading order in the transverse dynamics $\langle
\vec{I}\rangle=\hbar\operatorname{Re}g^{\uparrow\downarrow}\mathcal{S}/\left(
4\pi\right)  ,$ where the equal time and space correlation function
\begin{equation}
\mathcal{S=}\frac{1}{N_{c}}\sum_{i}^{N_{c}}\langle\dot{S}_{y,i}S_{x,i}-\dot
{S}_{x,i}S_{y,i}\rangle\label{eq:spin_pumping}%
\end{equation}
can be obtained either directly from the simulations or by integrating the
spectral power functions discussed above over frequency and momenta. At
equilibrium this current is exactly compensated by the thermal spin transfer
torque-induced currents from the metal. The observed spin Seebeck voltage then
reads, to leading order, $\Delta V_{SSE}\sim\mathcal{S}\partial T,$ where
$\partial T$ is the temperature gradient over the sample. We do not address
the proportionality constant here. The correlation functions that govern
electric spin injection, chemical potential-driven transport~\cite{Cornelissen:2015cz}, which might be relevant for the spin Seebeck effect~\cite{Cornelissen:2016tm}, as well as bulk magnonic transport that dominates the
signal for thick ferromagnets, are subject of ongoing study. In
Fig.~\ref{fig:spin_pumping} we plot $\mathcal{S}$ as a function of temperature
for YIG as well as for a hypothetical ferromagnet with parallel moments on a
simple BCC lattice (FM) that is tuned to the same saturation magnetization and
Curie temperature as YIG.

Both the ferrimagnet YIG and the FM show similar features. An increasing
temperature initially increases spin pumping by enhanced thermal agitation.
Close to the critical temperature spin pumping collapses to zero at
$T_{\mathrm{C}}$ together with the net magnetization. The YIG spin pumping is
maximized close to 300~K, i.e. far from the critical region of the phase
transition. This is caused by the increasing thermal occupation of the high
frequency optical mode with opposite chirality, which plays a significant role
above 300~K (see Fig.~\ref{fig:yig_optical}). A similar effect causes the low
temperature sign change of the spin Seebeck signal in
GdIG~\cite{Geprags:2016hh}. On the other hand, the thermal occupation of the
optical modes actually enhances the net magnetization rather than decreasing
it~\cite{Cherepanov:1993dd}.

Uchida et al. \cite{Uchida:2014jq} observed a power law $\Delta V_{SSE}%
\sim(T_{C}-T)^{3}$ close to the Curie temperature $T_{C},$ while the
magnetization scales $\sim(T_{C}-T)^{\tfrac{1}{2}}$ as expected from
mean-field theory. Here we find the same critical exponent $\tfrac{1}{2}$ for
both the magnetization and SSE effect, which can be rationalized in terms of
the spin wave gap that is closed in proportion with the exchange field. The
anomalous scaling found experimentally must be attributed to physical effects
not related to the dynamic susceptibility. A strong suppression of spin
transport in the ferrimagnet by large thermal fluctuations is a possible
explanation. On the other hand, the present spin Seebeck theory is based on
the Landau-Lifshitz-Gilbert equation which might not hold close to the phase
transition. More study is needed to understand the spin Seebeck effect close
to the critical regime. \ 

\textit{Conclusion} -- We present atomistic simulations of the spin dynamics
of the electrically insulating ferrimagnet yttrium iron garnet with
application to the spin Seebeck effect. The calculations transcend previous
theories by taking fully into account (i) the complicated crystal and
ferrimagnetic structure and (ii) the non-linearities caused by magnon-magnon
interactions at elevated temperatures. We observe a remarkable resilience of
the fundamental acoustic and optical modes with respect to thermal agitation,
which is explained by their large dispersion and spatial isolation from
numerous floppy modes with large heat capacity. At room temperature and above,
the ferrimagnetic optical mode is significantly occupied. Its negative
chirality leads to a suppression of thermal spin pumping and spin Seebeck
effect. The critical exponent observed for the spin Seebeck effect
\cite{Uchida:2014jq} remains as yet unexplained.

This work was supported by JSPS KAKENHI Grant Nos. 25247056, 25220910,
26103006 and the Tohoku University Graduate Program in Spintronics. We thank Jiang Xiao and Hiroto Adachi for useful discussions.

\bibliographystyle{apsrev4-1}
\bibliography{arXiv_yig_sse}

\begin{thebibliography}{26}%
\makeatletter
\providecommand \@ifxundefined [1]{%
 \@ifx{#1\undefined}
}%
\providecommand \@ifnum [1]{%
 \ifnum #1\expandafter \@firstoftwo
 \else \expandafter \@secondoftwo
 \fi
}%
\providecommand \@ifx [1]{%
 \ifx #1\expandafter \@firstoftwo
 \else \expandafter \@secondoftwo
 \fi
}%
\providecommand \natexlab [1]{#1}%
\providecommand \enquote  [1]{``#1''}%
\providecommand \bibnamefont  [1]{#1}%
\providecommand \bibfnamefont [1]{#1}%
\providecommand \citenamefont [1]{#1}%
\providecommand \href@noop [0]{\@secondoftwo}%
\providecommand \href [0]{\begingroup \@sanitize@url \@href}%
\providecommand \@href[1]{\@@startlink{#1}\@@href}%
\providecommand \@@href[1]{\endgroup#1\@@endlink}%
\providecommand \@sanitize@url [0]{\catcode `\\12\catcode `\$12\catcode
  `\&12\catcode `\#12\catcode `\^12\catcode `\_12\catcode `\%12\relax}%
\providecommand \@@startlink[1]{}%
\providecommand \@@endlink[0]{}%
\providecommand \url  [0]{\begingroup\@sanitize@url \@url }%
\providecommand \@url [1]{\endgroup\@href {#1}{\urlprefix }}%
\providecommand \urlprefix  [0]{URL }%
\providecommand \Eprint [0]{\href }%
\providecommand \doibase [0]{http://dx.doi.org/}%
\providecommand \selectlanguage [0]{\@gobble}%
\providecommand \bibinfo  [0]{\@secondoftwo}%
\providecommand \bibfield  [0]{\@secondoftwo}%
\providecommand \translation [1]{[#1]}%
\providecommand \BibitemOpen [0]{}%
\providecommand \bibitemStop [0]{}%
\providecommand \bibitemNoStop [0]{.\EOS\space}%
\providecommand \EOS [0]{\spacefactor3000\relax}%
\providecommand \BibitemShut  [1]{\csname bibitem#1\endcsname}%
\let\auto@bib@innerbib\@empty
\bibitem [{\citenamefont {Uchida}\ \emph {et~al.}(2010)\citenamefont {Uchida},
  \citenamefont {Xiao}, \citenamefont {Adachi}, \citenamefont {Ohe},
  \citenamefont {Takahashi}, \citenamefont {Ieda}, \citenamefont {Ota},
  \citenamefont {Kajiwara}, \citenamefont {Umezawa}, \citenamefont {Kawai},
  \citenamefont {Bauer}, \citenamefont {Maekawa},\ and\ \citenamefont
  {Saitoh}}]{Uchida:2010ei}%
  \BibitemOpen
  \bibfield  {author} {\bibinfo {author} {\bibfnamefont {K.}~\bibnamefont
  {Uchida}}, \bibinfo {author} {\bibfnamefont {J.}~\bibnamefont {Xiao}},
  \bibinfo {author} {\bibfnamefont {H.}~\bibnamefont {Adachi}}, \bibinfo
  {author} {\bibfnamefont {J.}~\bibnamefont {Ohe}}, \bibinfo {author}
  {\bibfnamefont {S.}~\bibnamefont {Takahashi}}, \bibinfo {author}
  {\bibfnamefont {J.}~\bibnamefont {Ieda}}, \bibinfo {author} {\bibfnamefont
  {T.}~\bibnamefont {Ota}}, \bibinfo {author} {\bibfnamefont {Y.}~\bibnamefont
  {Kajiwara}}, \bibinfo {author} {\bibfnamefont {H.}~\bibnamefont {Umezawa}},
  \bibinfo {author} {\bibfnamefont {H.}~\bibnamefont {Kawai}}, \bibinfo
  {author} {\bibfnamefont {G.~E.~W.}\ \bibnamefont {Bauer}}, \bibinfo {author}
  {\bibfnamefont {S.}~\bibnamefont {Maekawa}}, \ and\ \bibinfo {author}
  {\bibfnamefont {E.}~\bibnamefont {Saitoh}},\ }\href {\doibase
  10.1038/nmat2856} {\bibfield  {journal} {\bibinfo  {journal} {Nature Mater.}\
  }\textbf {\bibinfo {volume} {9}},\ \bibinfo {pages} {894} (\bibinfo {year}
  {2010})}\BibitemShut {NoStop}%
\bibitem [{\citenamefont {Bauer}\ \emph {et~al.}(2012)\citenamefont {Bauer},
  \citenamefont {Saitoh},\ and\ \citenamefont {van Wees}}]{Bauer:2012fq}%
  \BibitemOpen
  \bibfield  {author} {\bibinfo {author} {\bibfnamefont {G.~E.~W.}\
  \bibnamefont {Bauer}}, \bibinfo {author} {\bibfnamefont {E.}~\bibnamefont
  {Saitoh}}, \ and\ \bibinfo {author} {\bibfnamefont {B.~J.}\ \bibnamefont {van
  Wees}},\ }\href {\doibase 10.1038/nmat3301} {\bibfield  {journal} {\bibinfo
  {journal} {Nature Mater.}\ }\textbf {\bibinfo {volume} {11}},\ \bibinfo
  {pages} {391} (\bibinfo {year} {2012})}\BibitemShut {NoStop}%
\bibitem [{\citenamefont {Cherepanov}\ \emph {et~al.}(1993)\citenamefont
  {Cherepanov}, \citenamefont {Kolokolov},\ and\ \citenamefont
  {L'vov}}]{Cherepanov:1993dd}%
  \BibitemOpen
  \bibfield  {author} {\bibinfo {author} {\bibfnamefont {V.}~\bibnamefont
  {Cherepanov}}, \bibinfo {author} {\bibfnamefont {I.}~\bibnamefont
  {Kolokolov}}, \ and\ \bibinfo {author} {\bibfnamefont {V.}~\bibnamefont
  {L'vov}},\ }\href {\doibase 10.1016/0370-1573(93)90107-O} {\bibfield
  {journal} {\bibinfo  {journal} {Phys. Rep.}\ }\textbf {\bibinfo {volume}
  {229}},\ \bibinfo {pages} {81} (\bibinfo {year} {1993})}\BibitemShut
  {NoStop}%
\bibitem [{\citenamefont {Sun}\ \emph {et~al.}(2012)\citenamefont {Sun},
  \citenamefont {Song}, \citenamefont {Chang}, \citenamefont {Kabatek},
  \citenamefont {Jantz}, \citenamefont {Schneider}, \citenamefont {Wu},
  \citenamefont {Schultheiss},\ and\ \citenamefont {Hoffmann}}]{Sun:2012co}%
  \BibitemOpen
  \bibfield  {author} {\bibinfo {author} {\bibfnamefont {Y.}~\bibnamefont
  {Sun}}, \bibinfo {author} {\bibfnamefont {Y.-Y.}\ \bibnamefont {Song}},
  \bibinfo {author} {\bibfnamefont {H.}~\bibnamefont {Chang}}, \bibinfo
  {author} {\bibfnamefont {M.}~\bibnamefont {Kabatek}}, \bibinfo {author}
  {\bibfnamefont {M.}~\bibnamefont {Jantz}}, \bibinfo {author} {\bibfnamefont
  {W.}~\bibnamefont {Schneider}}, \bibinfo {author} {\bibfnamefont
  {M.}~\bibnamefont {Wu}}, \bibinfo {author} {\bibfnamefont {H.}~\bibnamefont
  {Schultheiss}}, \ and\ \bibinfo {author} {\bibfnamefont {A.}~\bibnamefont
  {Hoffmann}},\ }\href {\doibase 10.1063/1.4759039} {\bibfield  {journal}
  {\bibinfo  {journal} {Appl. Phys. Lett.}\ }\textbf {\bibinfo {volume}
  {101}},\ \bibinfo {pages} {152405} (\bibinfo {year} {2012})}\BibitemShut
  {NoStop}%
\bibitem [{\citenamefont {Xiao}\ \emph {et~al.}(2010)\citenamefont {Xiao},
  \citenamefont {Bauer}, \citenamefont {Uchida}, \citenamefont {Saitoh},\ and\
  \citenamefont {Maekawa}}]{Xiao:2010uc}%
  \BibitemOpen
  \bibfield  {author} {\bibinfo {author} {\bibfnamefont {J.}~\bibnamefont
  {Xiao}}, \bibinfo {author} {\bibfnamefont {G.~E.~W.}\ \bibnamefont {Bauer}},
  \bibinfo {author} {\bibfnamefont {K.-c.}\ \bibnamefont {Uchida}}, \bibinfo
  {author} {\bibfnamefont {E.}~\bibnamefont {Saitoh}}, \ and\ \bibinfo {author}
  {\bibfnamefont {S.}~\bibnamefont {Maekawa}},\ }\href {\doibase
  10.1103/PhysRevB.81.214418} {\bibfield  {journal} {\bibinfo  {journal} {Phys.
  Rev. B}\ }\textbf {\bibinfo {volume} {81}},\ \bibinfo {pages} {214418}
  (\bibinfo {year} {2010})}\BibitemShut {NoStop}%
\bibitem [{\citenamefont {Ritzmann}\ \emph {et~al.}(2014)\citenamefont
  {Ritzmann}, \citenamefont {Hinzke},\ and\ \citenamefont
  {Nowak}}]{Ritzmann:2014qn}%
  \BibitemOpen
  \bibfield  {author} {\bibinfo {author} {\bibfnamefont {U.}~\bibnamefont
  {Ritzmann}}, \bibinfo {author} {\bibfnamefont {D.}~\bibnamefont {Hinzke}}, \
  and\ \bibinfo {author} {\bibfnamefont {U.}~\bibnamefont {Nowak}},\ }\href
  {http://prb.aps.org/abstract/PRB/v89/i2/e024409} {\bibfield  {journal}
  {\bibinfo  {journal} {Phys. Rev. B}\ }\textbf {\bibinfo {volume} {89}},\
  \bibinfo {pages} {024409} (\bibinfo {year} {2014})}\BibitemShut {NoStop}%
\bibitem [{\citenamefont {Ohnuma}\ \emph {et~al.}(2013)\citenamefont {Ohnuma},
  \citenamefont {Adachi}, \citenamefont {Saitoh},\ and\ \citenamefont
  {Maekawa}}]{Ohnuma:2013il}%
  \BibitemOpen
  \bibfield  {author} {\bibinfo {author} {\bibfnamefont {Y.}~\bibnamefont
  {Ohnuma}}, \bibinfo {author} {\bibfnamefont {H.}~\bibnamefont {Adachi}},
  \bibinfo {author} {\bibfnamefont {E.}~\bibnamefont {Saitoh}}, \ and\ \bibinfo
  {author} {\bibfnamefont {S.}~\bibnamefont {Maekawa}},\ }\href {\doibase
  10.1103/PhysRevB.87.014423} {\bibfield  {journal} {\bibinfo  {journal} {Phys.
  Rev. B}\ }\textbf {\bibinfo {volume} {87}},\ \bibinfo {pages} {014423}
  (\bibinfo {year} {2013})}\BibitemShut {NoStop}%
\bibitem [{\citenamefont {Demokritov}\ \emph {et~al.}(2001)\citenamefont
  {Demokritov}, \citenamefont {Hillebrands},\ and\ \citenamefont
  {Slavin}}]{Demokritov:2001kk}%
  \BibitemOpen
  \bibfield  {author} {\bibinfo {author} {\bibfnamefont {S.~O.}\ \bibnamefont
  {Demokritov}}, \bibinfo {author} {\bibfnamefont {B.}~\bibnamefont
  {Hillebrands}}, \ and\ \bibinfo {author} {\bibfnamefont {A.~N.}\ \bibnamefont
  {Slavin}},\ }\href {\doibase 10.1016/S0370-1573(00)00116-2} {\bibfield
  {journal} {\bibinfo  {journal} {Phys. Rep.}\ }\textbf {\bibinfo {volume}
  {348}},\ \bibinfo {pages} {441} (\bibinfo {year} {2001})}\BibitemShut
  {NoStop}%
\bibitem [{\citenamefont {Plant}(1977)}]{Plant:1977tn}%
  \BibitemOpen
  \bibfield  {author} {\bibinfo {author} {\bibfnamefont {J.~S.}\ \bibnamefont
  {Plant}},\ }\href {http://iopscience.iop.org/0022-3719/10/23/014} {\bibfield
  {journal} {\bibinfo  {journal} {J. Phys. C: Solid State Phys.}\ }\textbf
  {\bibinfo {volume} {10}},\ \bibinfo {pages} {4805} (\bibinfo {year}
  {1977})}\BibitemShut {NoStop}%
\bibitem [{\citenamefont {Kikkawa}\ \emph {et~al.}(2015)\citenamefont
  {Kikkawa}, \citenamefont {Uchida}, \citenamefont {Daimon}, \citenamefont
  {Qiu}, \citenamefont {Shiomi},\ and\ \citenamefont
  {Saitoh}}]{Kikkawa:2015bn}%
  \BibitemOpen
  \bibfield  {author} {\bibinfo {author} {\bibfnamefont {T.}~\bibnamefont
  {Kikkawa}}, \bibinfo {author} {\bibfnamefont {K.-i.}\ \bibnamefont {Uchida}},
  \bibinfo {author} {\bibfnamefont {S.}~\bibnamefont {Daimon}}, \bibinfo
  {author} {\bibfnamefont {Z.}~\bibnamefont {Qiu}}, \bibinfo {author}
  {\bibfnamefont {Y.}~\bibnamefont {Shiomi}}, \ and\ \bibinfo {author}
  {\bibfnamefont {E.}~\bibnamefont {Saitoh}},\ }\href {\doibase
  10.1103/PhysRevB.92.064413} {\bibfield  {journal} {\bibinfo  {journal} {Phys.
  Rev. B}\ }\textbf {\bibinfo {volume} {92}},\ \bibinfo {pages} {064413}
  (\bibinfo {year} {2015})}\BibitemShut {NoStop}%
\bibitem [{\citenamefont {Jin}\ \emph {et~al.}(2015)\citenamefont {Jin},
  \citenamefont {Boona}, \citenamefont {Yang}, \citenamefont {Myers},\ and\
  \citenamefont {Heremans}}]{Jin:2015ik}%
  \BibitemOpen
  \bibfield  {author} {\bibinfo {author} {\bibfnamefont {H.}~\bibnamefont
  {Jin}}, \bibinfo {author} {\bibfnamefont {S.~R.}\ \bibnamefont {Boona}},
  \bibinfo {author} {\bibfnamefont {Z.}~\bibnamefont {Yang}}, \bibinfo {author}
  {\bibfnamefont {R.~C.}\ \bibnamefont {Myers}}, \ and\ \bibinfo {author}
  {\bibfnamefont {J.~P.}\ \bibnamefont {Heremans}},\ }\href {\doibase
  10.1103/PhysRevB.92.054436} {\bibfield  {journal} {\bibinfo  {journal} {Phys.
  Rev. B}\ }\textbf {\bibinfo {volume} {92}},\ \bibinfo {pages} {054436}
  (\bibinfo {year} {2015})}\BibitemShut {NoStop}%
\bibitem [{\citenamefont {Guo}\ \emph {et~al.}(2015)\citenamefont {Guo},
  \citenamefont {Kehlberger}, \citenamefont {Cramer}, \citenamefont {Jakob},\
  and\ \citenamefont {Kl{\"a}ui}}]{Guo:2015vi}%
  \BibitemOpen
  \bibfield  {author} {\bibinfo {author} {\bibfnamefont {E.-J.}\ \bibnamefont
  {Guo}}, \bibinfo {author} {\bibfnamefont {A.}~\bibnamefont {Kehlberger}},
  \bibinfo {author} {\bibfnamefont {J.}~\bibnamefont {Cramer}}, \bibinfo
  {author} {\bibfnamefont {G.}~\bibnamefont {Jakob}}, \ and\ \bibinfo {author}
  {\bibfnamefont {M.}~\bibnamefont {Kl{\"a}ui}},\ }\href
  {http://arxiv.org/abs/1506.06037v1} {\bibfield  {journal} {\bibinfo
  {journal} {arXiv}\ } (\bibinfo {year} {2015})},\ \Eprint
  {http://arxiv.org/abs/1506.06037v1} {1506.06037v1} \BibitemShut {NoStop}%
\bibitem [{\citenamefont {Cornelissen}\ \emph {et~al.}(2015)\citenamefont
  {Cornelissen}, \citenamefont {Liu}, \citenamefont {Duine}, \citenamefont
  {Youssef},\ and\ \citenamefont {van Wees}}]{Cornelissen:2015cz}%
  \BibitemOpen
  \bibfield  {author} {\bibinfo {author} {\bibfnamefont {L.~J.}\ \bibnamefont
  {Cornelissen}}, \bibinfo {author} {\bibfnamefont {J.}~\bibnamefont {Liu}},
  \bibinfo {author} {\bibfnamefont {R.~A.}\ \bibnamefont {Duine}}, \bibinfo
  {author} {\bibfnamefont {J.~B.}\ \bibnamefont {Youssef}}, \ and\ \bibinfo
  {author} {\bibfnamefont {B.~J.}\ \bibnamefont {van Wees}},\ }\href {\doibase
  10.1038/nphys3465} {\bibfield  {journal} {\bibinfo  {journal} {Nat Phys}\
  }\textbf {\bibinfo {volume} {11}},\ \bibinfo {pages} {1022} (\bibinfo {year}
  {2015})}\BibitemShut {NoStop}%
\bibitem [{\citenamefont {Gepraegs}\ \emph {et~al.}(2016)\citenamefont
  {Gepraegs}, \citenamefont {Kehlberger}, \citenamefont {Della~Coletta},
  \citenamefont {Qiu}, \citenamefont {Guo}, \citenamefont {Schulz},
  \citenamefont {Mix}, \citenamefont {Meyer}, \citenamefont {Kamra},
  \citenamefont {Althammer}, \citenamefont {Huebl}, \citenamefont {Jakob},
  \citenamefont {Ohnuma}, \citenamefont {Adachi}, \citenamefont {Barker},
  \citenamefont {Maekawa}, \citenamefont {Bauer}, \citenamefont {Saitoh},
  \citenamefont {Gross}, \citenamefont {Goennenwein},\ and\ \citenamefont
  {Klaeui}}]{Geprags:2016hh}%
  \BibitemOpen
  \bibfield  {author} {\bibinfo {author} {\bibfnamefont {S.}~\bibnamefont
  {Gepraegs}}, \bibinfo {author} {\bibfnamefont {A.}~\bibnamefont
  {Kehlberger}}, \bibinfo {author} {\bibfnamefont {F.}~\bibnamefont
  {Della~Coletta}}, \bibinfo {author} {\bibfnamefont {Z.}~\bibnamefont {Qiu}},
  \bibinfo {author} {\bibfnamefont {E.-J.}\ \bibnamefont {Guo}}, \bibinfo
  {author} {\bibfnamefont {T.}~\bibnamefont {Schulz}}, \bibinfo {author}
  {\bibfnamefont {C.}~\bibnamefont {Mix}}, \bibinfo {author} {\bibfnamefont
  {S.}~\bibnamefont {Meyer}}, \bibinfo {author} {\bibfnamefont
  {A.}~\bibnamefont {Kamra}}, \bibinfo {author} {\bibfnamefont
  {M.}~\bibnamefont {Althammer}}, \bibinfo {author} {\bibfnamefont
  {H.}~\bibnamefont {Huebl}}, \bibinfo {author} {\bibfnamefont
  {G.}~\bibnamefont {Jakob}}, \bibinfo {author} {\bibfnamefont
  {Y.}~\bibnamefont {Ohnuma}}, \bibinfo {author} {\bibfnamefont
  {H.}~\bibnamefont {Adachi}}, \bibinfo {author} {\bibfnamefont
  {J.}~\bibnamefont {Barker}}, \bibinfo {author} {\bibfnamefont
  {S.}~\bibnamefont {Maekawa}}, \bibinfo {author} {\bibfnamefont {G.~E.~W.}\
  \bibnamefont {Bauer}}, \bibinfo {author} {\bibfnamefont {E.}~\bibnamefont
  {Saitoh}}, \bibinfo {author} {\bibfnamefont {R.}~\bibnamefont {Gross}},
  \bibinfo {author} {\bibfnamefont {S.~T.~B.}\ \bibnamefont {Goennenwein}}, \
  and\ \bibinfo {author} {\bibfnamefont {M.}~\bibnamefont {Klaeui}},\ }\href
  {\doibase 10.1038/ncomms10452} {\bibfield  {journal} {\bibinfo  {journal}
  {Nat. Commun.}\ }\textbf {\bibinfo {volume} {7}},\ \bibinfo {pages} {10452}
  (\bibinfo {year} {2016})}\BibitemShut {NoStop}%
\bibitem [{\citenamefont {Uchida}\ \emph
  {et~al.}(2014{\natexlab{a}})\citenamefont {Uchida}, \citenamefont {Ishida},
  \citenamefont {Kikkawa}, \citenamefont {Kirihara}, \citenamefont {Murakami},\
  and\ \citenamefont {Saitoh}}]{Uchida:2014bj}%
  \BibitemOpen
  \bibfield  {author} {\bibinfo {author} {\bibfnamefont {K.}~\bibnamefont
  {Uchida}}, \bibinfo {author} {\bibfnamefont {M.}~\bibnamefont {Ishida}},
  \bibinfo {author} {\bibfnamefont {T.}~\bibnamefont {Kikkawa}}, \bibinfo
  {author} {\bibfnamefont {A.}~\bibnamefont {Kirihara}}, \bibinfo {author}
  {\bibfnamefont {T.}~\bibnamefont {Murakami}}, \ and\ \bibinfo {author}
  {\bibfnamefont {E.}~\bibnamefont {Saitoh}},\ }\href {\doibase
  10.1088/0953-8984/26/38/389601} {\bibfield  {journal} {\bibinfo  {journal}
  {J. Phys.: Condens. Matter}\ }\textbf {\bibinfo {volume} {26}},\ \bibinfo
  {pages} {389601} (\bibinfo {year} {2014}{\natexlab{a}})}\BibitemShut
  {NoStop}%
\bibitem [{\citenamefont {Uchida}\ \emph {et~al.}(2016)\citenamefont {Uchida},
  \citenamefont {Adachi}, \citenamefont {Kikkawa}, \citenamefont {Kirihara},
  \citenamefont {Ishida}, \citenamefont {Yorozu}, \citenamefont {Maekawa},\
  and\ \citenamefont {Saitoh}}]{Uchida:2016jo}%
  \BibitemOpen
  \bibfield  {author} {\bibinfo {author} {\bibfnamefont {K.}~\bibnamefont
  {Uchida}}, \bibinfo {author} {\bibfnamefont {H.}~\bibnamefont {Adachi}},
  \bibinfo {author} {\bibfnamefont {T.}~\bibnamefont {Kikkawa}}, \bibinfo
  {author} {\bibfnamefont {A.}~\bibnamefont {Kirihara}}, \bibinfo {author}
  {\bibfnamefont {M.}~\bibnamefont {Ishida}}, \bibinfo {author} {\bibfnamefont
  {S.}~\bibnamefont {Yorozu}}, \bibinfo {author} {\bibfnamefont
  {S.}~\bibnamefont {Maekawa}}, \ and\ \bibinfo {author} {\bibfnamefont
  {E.}~\bibnamefont {Saitoh}},\ }\href {\doibase 10.1109/JPROC.2016.2535167}
  {\bibfield  {journal} {\bibinfo  {journal} {Proc. IEEE}\ }\textbf {\bibinfo
  {volume} {PP}},\ \bibinfo {pages} {1} (\bibinfo {year} {2016})}\BibitemShut
  {NoStop}%
\bibitem [{\citenamefont {Hinzke}\ and\ \citenamefont
  {Nowak}(1999)}]{Hinzke:1999ud}%
  \BibitemOpen
  \bibfield  {author} {\bibinfo {author} {\bibfnamefont {D.}~\bibnamefont
  {Hinzke}}\ and\ \bibinfo {author} {\bibfnamefont {U.}~\bibnamefont {Nowak}},\
  }\href@noop {} {\bibfield  {journal} {\bibinfo  {journal} {Comput. Phys.
  Commun.}\ }\textbf {\bibinfo {volume} {121}},\ \bibinfo {pages} {334}
  (\bibinfo {year} {1999})}\BibitemShut {NoStop}%
\bibitem [{\citenamefont {Anderson}(1964)}]{Anderson:1964kv}%
  \BibitemOpen
  \bibfield  {author} {\bibinfo {author} {\bibfnamefont {E.~E.}\ \bibnamefont
  {Anderson}},\ }\href {\doibase 10.1103/PhysRev.134.A1581} {\bibfield
  {journal} {\bibinfo  {journal} {Phys. Rev.}\ }\textbf {\bibinfo {volume}
  {134}},\ \bibinfo {pages} {A1581} (\bibinfo {year} {1964})}\BibitemShut
  {NoStop}%
\bibitem [{\citenamefont {Oitmaa}\ and\ \citenamefont
  {Falk}(2009)}]{Oitmaa:2009jm}%
  \BibitemOpen
  \bibfield  {author} {\bibinfo {author} {\bibfnamefont {J.}~\bibnamefont
  {Oitmaa}}\ and\ \bibinfo {author} {\bibfnamefont {T.}~\bibnamefont {Falk}},\
  }\href {\doibase 10.1088/0953-8984/21/12/124212} {\bibfield  {journal}
  {\bibinfo  {journal} {J. Phys.: Condens. Matter}\ }\textbf {\bibinfo {volume}
  {21}},\ \bibinfo {pages} {124212} (\bibinfo {year} {2009})}\BibitemShut
  {NoStop}%
\bibitem [{\citenamefont {Wangsness}(1953)}]{Wangsness:1953vga}%
  \BibitemOpen
  \bibfield  {author} {\bibinfo {author} {\bibfnamefont {R.~K.}\ \bibnamefont
  {Wangsness}},\ }\href {http://prola.aps.org/abstract/PR/v91/i5/p1085_1}
  {\bibfield  {journal} {\bibinfo  {journal} {Phys. Rev.}\ }\textbf {\bibinfo
  {volume} {91}},\ \bibinfo {pages} {1085} (\bibinfo {year}
  {1953})}\BibitemShut {NoStop}%
\bibitem [{\citenamefont {Harris}(1963)}]{Harris:1963vs}%
  \BibitemOpen
  \bibfield  {author} {\bibinfo {author} {\bibfnamefont {A.~B.}\ \bibnamefont
  {Harris}},\ }\href {http://prola.aps.org/abstract/PR/v132/i6/p2398_1}
  {\bibfield  {journal} {\bibinfo  {journal} {Phys. Rev.}\ }\textbf {\bibinfo
  {volume} {132}},\ \bibinfo {pages} {2398} (\bibinfo {year}
  {1963})}\BibitemShut {NoStop}%
\bibitem [{\citenamefont {LeCraw}\ and\ \citenamefont
  {Walker}(1961)}]{LeCraw:1961ef}%
  \BibitemOpen
  \bibfield  {author} {\bibinfo {author} {\bibfnamefont {R.~C.}\ \bibnamefont
  {LeCraw}}\ and\ \bibinfo {author} {\bibfnamefont {L.~R.}\ \bibnamefont
  {Walker}},\ }\href {\doibase 10.1063/1.2000390} {\bibfield  {journal}
  {\bibinfo  {journal} {J. Appl. Phys.}\ }\textbf {\bibinfo {volume} {32}},\
  \bibinfo {pages} {S167} (\bibinfo {year} {1961})}\BibitemShut {NoStop}%
\bibitem [{\citenamefont {Bastardis}\ \emph {et~al.}(2012)\citenamefont
  {Bastardis}, \citenamefont {Atxitia}, \citenamefont {Chubykalo-Fesenko},\
  and\ \citenamefont {Kachkachi}}]{Bastardis:2012cs}%
  \BibitemOpen
  \bibfield  {author} {\bibinfo {author} {\bibfnamefont {R.}~\bibnamefont
  {Bastardis}}, \bibinfo {author} {\bibfnamefont {U.}~\bibnamefont {Atxitia}},
  \bibinfo {author} {\bibfnamefont {O.}~\bibnamefont {Chubykalo-Fesenko}}, \
  and\ \bibinfo {author} {\bibfnamefont {H.}~\bibnamefont {Kachkachi}},\ }\href
  {\doibase 10.1103/PhysRevB.86.094415} {\bibfield  {journal} {\bibinfo
  {journal} {Phys. Rev. B}\ }\textbf {\bibinfo {volume} {86}},\ \bibinfo
  {pages} {094415} (\bibinfo {year} {2012})}\BibitemShut {NoStop}%
\bibitem [{\citenamefont {Jia}\ \emph {et~al.}(2011)\citenamefont {Jia},
  \citenamefont {Liu}, \citenamefont {Xia},\ and\ \citenamefont
  {Bauer}}]{Jia:2011gm}%
  \BibitemOpen
  \bibfield  {author} {\bibinfo {author} {\bibfnamefont {X.}~\bibnamefont
  {Jia}}, \bibinfo {author} {\bibfnamefont {K.}~\bibnamefont {Liu}}, \bibinfo
  {author} {\bibfnamefont {K.}~\bibnamefont {Xia}}, \ and\ \bibinfo {author}
  {\bibfnamefont {G.~E.~W.}\ \bibnamefont {Bauer}},\ }\href {\doibase
  10.1209/0295-5075/96/17005} {\bibfield  {journal} {\bibinfo  {journal}
  {Europhys. Lett.}\ }\textbf {\bibinfo {volume} {96}},\ \bibinfo {pages}
  {17005} (\bibinfo {year} {2011})}\BibitemShut {NoStop}%
\bibitem [{\citenamefont {Cornelissen}\ \emph {et~al.}(2016)\citenamefont
  {Cornelissen}, \citenamefont {Peters}, \citenamefont {Duine}, \citenamefont
  {Bauer},\ and\ \citenamefont {van Wees}}]{Cornelissen:2016tm}%
  \BibitemOpen
  \bibfield  {author} {\bibinfo {author} {\bibfnamefont {L.~J.}\ \bibnamefont
  {Cornelissen}}, \bibinfo {author} {\bibfnamefont {K.~J.~H.}\ \bibnamefont
  {Peters}}, \bibinfo {author} {\bibfnamefont {R.~A.}\ \bibnamefont {Duine}},
  \bibinfo {author} {\bibfnamefont {G.~E.~W.}\ \bibnamefont {Bauer}}, \ and\
  \bibinfo {author} {\bibfnamefont {B.~J.}\ \bibnamefont {van Wees}},\ }\href
  {http://arxiv.org/abs/1604.03706} {\bibfield  {journal} {\bibinfo  {journal}
  {arXiv}\ } (\bibinfo {year} {2016})},\ \Eprint
  {http://arxiv.org/abs/1604.03706} {1604.03706} \BibitemShut {NoStop}%
\bibitem [{\citenamefont {Uchida}\ \emph
  {et~al.}(2014{\natexlab{b}})\citenamefont {Uchida}, \citenamefont {Kikkawa},
  \citenamefont {Miura}, \citenamefont {Shiomi},\ and\ \citenamefont
  {Saitoh}}]{Uchida:2014jq}%
  \BibitemOpen
  \bibfield  {author} {\bibinfo {author} {\bibfnamefont {K.-i.}\ \bibnamefont
  {Uchida}}, \bibinfo {author} {\bibfnamefont {T.}~\bibnamefont {Kikkawa}},
  \bibinfo {author} {\bibfnamefont {A.}~\bibnamefont {Miura}}, \bibinfo
  {author} {\bibfnamefont {J.}~\bibnamefont {Shiomi}}, \ and\ \bibinfo {author}
  {\bibfnamefont {E.}~\bibnamefont {Saitoh}},\ }\href {\doibase
  10.1103/PhysRevX.4.041023} {\bibfield  {journal} {\bibinfo  {journal} {Phys.
  Rev. X}\ }\textbf {\bibinfo {volume} {4}},\ \bibinfo {pages} {041023}
  (\bibinfo {year} {2014}{\natexlab{b}})}\BibitemShut {NoStop}%
\end{thebibliography}%

\end{document}